\providecommand{\reserveinserts}[1]{}

\let\MRMOriginalBibliographyStyle\bibliographystyle

\renewcommand{\bibliographystyle}[1]{\MRMOriginalBibliographyStyle{MRM-AMA-compatible}}

\documentclass[AMA,STIX2COL,Linenumbersoff]{MRM}

\usepackage{array,booktabs,colortbl,placeins,xurl}

\hypersetup{linkcolor=black}

\definecolor{TableTint}{RGB}{239,244,247}

\definecolor{TableAccent}{RGB}{40,103,116}

\articletype{}

\historydates{}

\doiheadtext{}

\renewcommand{\printjnlcitation}{}

\hypersetup{pdftitle={MRFoundry: Knowledge-guided agent loops for end-to-end MRI research},pdfauthor={Ke Wang; Atakan Topcu; Hossein Momeni; Molin Zhang}}

\begin{document}

\title{MRFoundry: Knowledge-guided agent loops for end-to-end MRI research}

\author[1]{Ke Wang}{}

\author[2]{Atakan Topcu}{}

\author[2]{Hossein Momeni}{}

\author[1]{Molin Zhang}{}

\authormark{WANG ET AL.}

\address[1]{\orgname{Independent Researcher}}

\address[2]{\orgdiv{Department of Electrical Engineering and Computer Sciences}, \orgname{University of California, Berkeley}, Berkeley, California, USA}

\corres{Ke Wang. \email{kewang0622@gmail.com}}

\abstract[Abstract]{AI agents can write code, use scientific software, and revise computational workflows, opening new possibilities for MRI research. Effective assistance requires connecting acquisition physics, data conventions, specialized tools, and scientific validation. We present MRFoundry, an open-source, extensible package that connects MRI knowledge to agent actions and output checks. Project-local memory preserves decisions and evidence, while reviewed contributions allow researchers to improve shared guidance. Researcher-supervised examples demonstrate skill-guided sequence auditing and acquired-data reconstruction, connecting tool use, output checks, and documented decisions. Alongside MRFoundry, we introduce MRI Agent Bench to compare agents with and without the package across models and MRI task domains. Across 240 paired Core evaluations spanning 24 task types and ten models, all-checks-passed completion increased from 67.5\% without MRFoundry to 78.3\% with it. Across 134 Hard pairs, completion increased from 36.6\% to 44.8\%. MRFoundry makes MRI workflow expertise available as reusable guidance that researchers can inspect, adapt, and extend within existing agent and scientific software environments. MRFoundry is openly available at \url{https://github.com/KeWang0622/MRFoundry}.}


\maketitle

\section{Introduction}

AI agents can write code, operate software, and use execution feedback to carry out computational work \cite{react,claudeloop,sweagent,datainterpreter}. In scientific workflows, these capabilities offer a way to connect specialized tools, inspect intermediate results, and revise an implementation as a research question develops \cite{coscientist}. MRI is a natural setting for this assistance: acquisition design, simulation, reconstruction, analysis, and reporting span several software environments while sharing assumptions about the same experiment.

Effective assistance requires knowledge of acquisition physics, hardware constraints, data conventions, and computational models \cite{bernstein}. A reconstruction can execute successfully with incorrect axes or sampling conventions; a valid sequence file still requires physical and numerical checks. The agent must establish what the inputs represent, preserve their meaning between applications, and select evidence that supports interpretation. For a sequence prototype, this means connecting requested timing to an exported sequence and a simulation that reads it correctly. For reconstruction, it means relating measured k-space to the sampling mask, encoding model, and comparison image. The knowledge needed for these decisions is distributed across publications, documentation, and laboratory experience. Making it accessible during a task is the opportunity addressed here.

We developed MRFoundry as an extensible package of MRI guidance for existing agents \cite{mrfoundry}. Each skill links scientific references and software instructions to actions and output checks. A project-local notebook preserves the evidence behind decisions, and selected lessons can inform shared guidance through human review. The researcher supplies the question and constraints, and the agent performs computational work using established applications and supplied helper scripts. Related systems address other parts of this process: Agent4MR couples sequence generation to physics-aware validation, and BCER supplies a controller for long MRI workflows with explicit artifact binding \cite{agent4mr,bcer}. MRFoundry provides reusable guidance across MRI domains, while execution and recovery remain the host's responsibility.

This paper describes the toolbox, project memory, and contribution workflow, followed by two researcher-supervised examples. The sequence example includes a later skill-guided audit; the reconstruction example documents guidance consulted during execution. Alongside MRFoundry, we introduce MRI Agent Bench to compare matched agents with and without the toolbox. Its specifications and public tasks are maintained in a separate repository \cite{blochbenchtasks,blochbench}. Together, the demonstrations and paired comparison connect the package design to concrete research decisions and measured task completion.

\section{Methods}

We first describe how the toolbox supplies guidance, retains project evidence, and supports reviewed contributions. We then specify the supervised demonstrations and the paired benchmark comparison, which address workflow behavior and task completion, respectively.

\subsection{Knowledge-guided research workflows}

\begin{figure*}[!tp]
\centering
\includegraphics[width=0.96\textwidth,height=0.80\textheight,keepaspectratio]{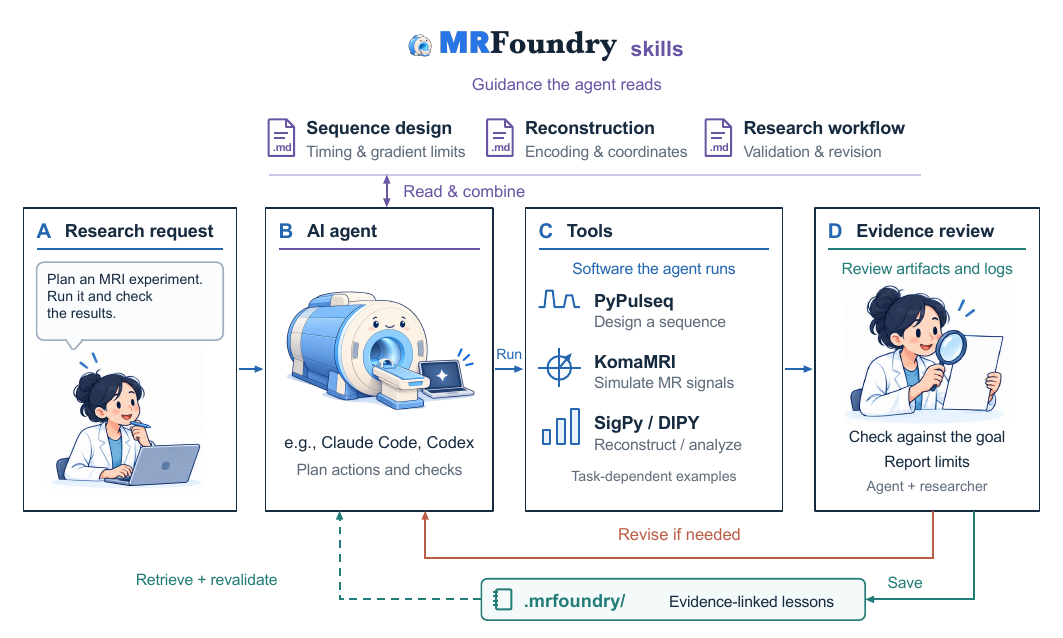}
\caption{MRFoundry's guidance-based research loop \cite{mrfoundry}. A: the researcher specifies a goal. B: an AI agent consults relevant skills and plans actions and checks; Claude Code and Codex are illustrative hosts \cite{claudecode,codex}. C: separately installed applications perform task-dependent computations, illustrated by PyPulseq, KomaMRI, SigPy and DIPY \cite{pypulseq,koma,sigpy,dipy}. D: artifacts are reviewed with the researcher against the goal; insufficient evidence can prompt revision. The project-memory guide prescribes evidence-linked records, selective retrieval and revalidation \cite{memoryguide}. Arrows describe recommended relationships rather than an enforced execution policy. Generated characters identify roles and are not scientific data. MRFoundry is compared with no skills in \hyperref[tab:2]{Table~2}; the effect of project memory on later tasks was not measured.}
\label{fig:1}
\end{figure*}

\hyperref[fig:1]{Figure~1} shows the intended loop. The researcher supplies a question, inputs, and constraints. The agent identifies relevant skills, plans actions, executes tools, and checks whether the outputs address the question. Failed checks can prompt revision of an implementation or its assumptions. MRFoundry describes this process in instructions; the host and researcher control execution and decide when further work is needed.

\subsubsection{Skill organization and MRI knowledge}

Each skill starts with a description of when to use it and links to detailed instructions, references, and helper scripts where needed. Instructions guide decisions; linked references supply implementation details. The host agent interprets this material in the context of the request, so the same skill can guide tasks with different data and constraints. This organization separates reusable MRI guidance from the particular commands and numerical applications used in a project.

\hyperref[tab:1]{Table~1} summarizes the initial collection of general and specialist skills, providing a starting point for community extension. The MRI navigator connects concepts, formats, software, and references across the field. The research-workflow skill connects study design, experiments, evaluation, and writing \cite{mrfoundry}. Specialist skills cover hardware, pulse-sequence design, conventional reconstruction, deep-learning reconstruction, and diffusion MRI. They address decisions such as choosing an encoding model, checking gradient conventions, or establishing reconstruction baselines. Skills can be combined when a task crosses domains: sequence design may require hardware limits and reconstruction conventions, while a learned reconstruction study requires conventional comparisons and a defensible evaluation design.

\begin{table*}[!tp]
\caption{MRI-domain skills and reference entry points}
\label{tab:1}
\begingroup
\tablebodyfont
\setlength{\tabcolsep}{5pt}
\renewcommand{\arraystretch}{1.0}
\noindent\begin{tabular}{@{}>{\raggedright\arraybackslash}p{0.24\dimexpr\textwidth-20pt\relax}>{\raggedright\arraybackslash}p{0.31\dimexpr\textwidth-20pt\relax}>{\raggedright\arraybackslash}p{0.45\dimexpr\textwidth-20pt\relax}@{}}
\toprule
\rowcolor{TableTint}\textbf{MRI skill} & \textbf{Decisions supported} & \textbf{Reference entry points}\\
\midrule\addlinespace[3pt]
\textbf{MRI navigator}\newline {\tablebodyfont\selectfont \href{https://github.com/KeWang0622/MRFoundry/blob/main/skills/mri-research/SKILL.md}{\texttt{mri-\allowbreak research}}} & Physics, data conventions and cross-domain navigation & MRI and sequence foundations \cite{bernstein}\\
\addlinespace[3pt]
\textbf{Research workflow}\newline {\tablebodyfont\selectfont \href{https://github.com/KeWang0622/MRFoundry/blob/main/skills/mri-research-workflow/SKILL.md}{\texttt{mri-\allowbreak research-\allowbreak workflow}}} & Study design, evidence review and reporting & Reproducible neuroimaging; AI reporting \cite{cobidas,claim}\\
\addlinespace[3pt]
\textbf{Hardware}\newline {\tablebodyfont\selectfont \href{https://github.com/KeWang0622/MRFoundry/blob/main/skills/mri-hardware/SKILL.md}{\texttt{mri-\allowbreak hardware}}} & Receive weighting and acquisition constraints & Receive-array foundations \cite{roemer}\\
\addlinespace[3pt]
\textbf{Sequence design}\newline {\tablebodyfont\selectfont \href{https://github.com/KeWang0622/MRFoundry/blob/main/skills/pulse-sequence-design/SKILL.md}{\texttt{pulse-\allowbreak sequence-\allowbreak design}}} & Events, timing, trajectories and simulation & Pulseq, PyPulseq and KomaMRI \cite{pulseq,pypulseq,koma}\\
\addlinespace[3pt]
\textbf{Reconstruction}\newline {\tablebodyfont\selectfont \href{https://github.com/KeWang0622/MRFoundry/blob/main/skills/mri-reconstruction/SKILL.md}{\texttt{mri-\allowbreak reconstruction}}} & Encoding, calibration and regularization & SENSE, ESPIRiT and sparse MRI \cite{sense,espirit,sparsemri}; SigPy and BART \cite{sigpy,bart}\\
\addlinespace[3pt]
\textbf{Deep-learning reconstruction}\newline {\tablebodyfont\selectfont \href{https://github.com/KeWang0622/MRFoundry/blob/main/skills/deep-learning-recon/SKILL.md}{\texttt{deep-\allowbreak learning-\allowbreak recon}}} & Training, splits, baselines and generalization & fastMRI and end-to-end variational networks \cite{fastmri,varnet}\\
\addlinespace[3pt]
\textbf{Diffusion MRI}\newline {\tablebodyfont\selectfont \href{https://github.com/KeWang0622/MRFoundry/blob/main/skills/diffusion-mri/SKILL.md}{\texttt{diffusion-\allowbreak mri}}} & Gradients, model assumptions and fit interpretation & Diffusion tensor, DIPY and WLS \cite{basser,dipy,chungwls}\\
\addlinespace[3pt]\bottomrule\end{tabular}
\par\vspace{6pt}\tablefootnotefont The public source index links each stable skill name to its annotated references \cite{mrfoundryindex}. These are representative entry points. Skills provide guidance and helper scripts; broader software dependencies are installed separately. Topic coverage is not an evaluation result.
\endgroup
\end{table*}

The reference index connects instructions to annotated papers, textbooks, courses, and software documentation \cite{mrfoundryindex}. Scientific sources explain the basis for a method, while software documentation describes its implementation. For example, a reconstruction task can begin with the specialist skill, follow a format reference to establish array conventions, and consult workflow guidance to define a baseline and endpoint. The agent and researcher can thus trace a recommended action to its source. \hyperref[tab:1]{Table~1} describes intended coverage; the demonstrations illustrate selected workflows, and the benchmark assesses specified task completion.

\subsubsection{Agent workflow and tool use}

Hosts such as Claude Code and Codex expose the skills to the agent \cite{claudecode,codex}. The command \texttt{npx skills add KeWang0622/\allowbreak MRFoundry} installs the guidance; \texttt{-\allowbreak -\allowbreak skill mri-\allowbreak reconstruction} selects one skill \cite{skillscli,mrfoundry}. Scientific applications and restricted datasets retain their installation, access, and license requirements. MRFoundry combines guidance with task-specific helper scripts; broader software dependencies are installed separately.

Before execution, guidance asks the agent to inspect the input data and working environment, resolve missing information, install dependencies in isolation, and check a small upstream example \cite{toolsetup}. When connecting applications, the agent must preserve units, axes, sample order, and model assumptions. For example, reconstruction requires distinguishing the stored grid from acquired support before constructing a mask. After execution, the agent inspects artifacts and quantitative checks against the research request. To interpret the result, the researcher also needs to know how the inputs were understood, what served as the comparison reference, and which questions remain unresolved.

\subsection{Research memory and reusable lessons}

\begin{figure*}[!tp]
\centering
\includegraphics[width=0.94\textwidth,height=0.80\textheight,keepaspectratio]{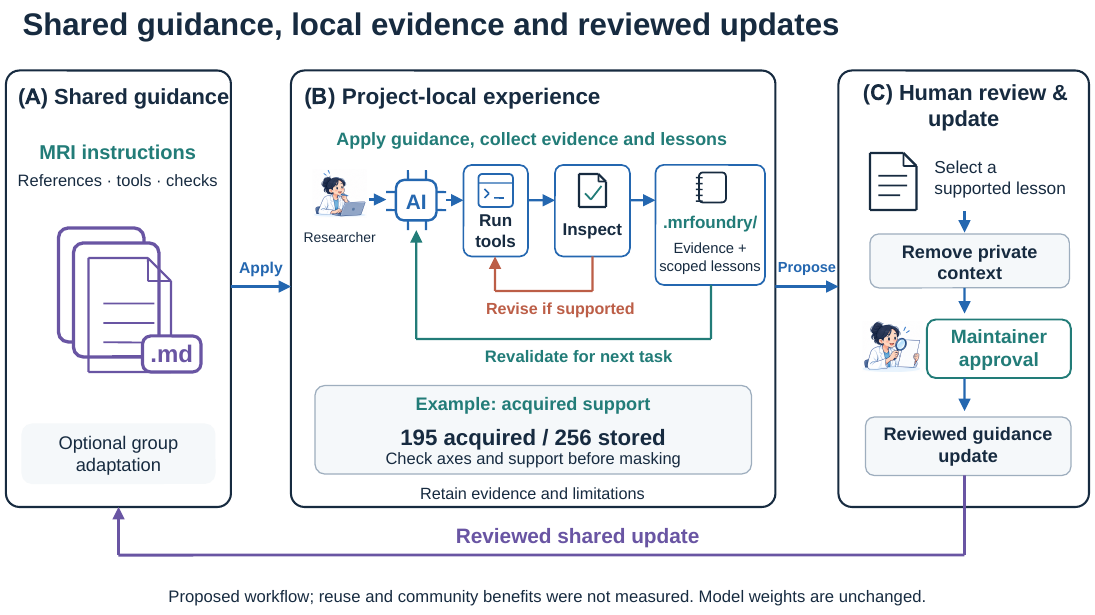}
\caption{Shared guidance, local experience and reviewed improvement. A: Public MRFoundry skills provide reusable MRI instructions and source links; a group can adapt these instructions to its methods and environment. B: Project-local records retain preferences, evidence and scoped lessons. The researcher and agent execute tasks, inspect outputs and record evidence; current work can be revised when supported, while retrieved lessons must be revalidated for a new task. The acquired-support example distinguishes 195 acquired from 256 stored phase-encoding columns in the inspected source; the value is not a general acquisition rule. C: Researchers select supported changes, remove private context and supply evidence for maintainer review before shared guidance is updated. Arrows describe the proposed guidance workflow rather than an enforced execution policy. This improvement loop was not evaluated in the fixed-snapshot Core/Hard comparison; model weights are unchanged.}
\label{fig:2}
\end{figure*}

The workflow also produces information that is specific to the project. \hyperref[fig:2]{Figure~2} shows how the \texttt{.mrfoundry/\allowbreak } notebook retains this evidence separately from reusable public skills. It contains an index, working preferences, environment information, evidence-linked lessons, and individual run records \cite{memoryguide}. Run records retain requests, input identifiers, versions, parameters, commands, outputs, and unsuccessful attempts. Together, these details allow a later session to recover both a decision and the reasoning behind it.

A reusable lesson states an observation, supporting evidence, scope, and revalidation conditions. Retrieval begins with the index: the agent selects lessons and checks their applicability to the current data and environment. Conflicting evidence can narrow or supersede a lesson while preserving its original record. Preferences, such as reporting format, are stored separately so that local working habits do not become methodological rules.

This recursive self-improvement design updates written guidance rather than model weights. The initializer can add retrieval instructions to project \texttt{AGENTS.md} or \texttt{CLAUDE.md} files; the host controls their loading and execution \cite{memoryguide}. Commands may depend on a software version or dataset, so reuse requires revalidation. Evidence-linked recording helps researchers decide which parts of a previous workflow apply to the next request.

\subsection{Local adaptation and public contribution}

Project memory also provides a starting point for improving shared guidance. Researchers can keep a lesson local, adapt instructions to group conventions, or propose a generalizable change to the public toolbox. Project notebooks are excluded from Git by default. Preparing a public contribution requires selecting the relevant instruction, removing private context, and supplying a scientific source or tested example \cite{memoryguide,contributing}. The notebook itself is not automatically published.

Local adaptation should distinguish working conventions from scientific assumptions. A group can adopt a preferred folder structure, whereas a sampling rule requires justification for the current data. The project record retains that context for revalidation. Public contributions should explain the problem addressed, applicable conditions, and evidence supporting the recommendation \cite{memoryguide,contributing}.

Contributors edit a skill or reference on a branch and open a pull request (PR). Automated checks validate structure and links; a maintainer reviews scientific relevance and scope before merging \cite{contributing}. For example, dataset-specific dimensions can motivate a general instruction to verify the release format, while those dimensions remain local. Groups review updates before adoption. This process turns project lessons into inspectable, source-supported instructions for other researchers.

\subsection{Demonstration design}

\begin{figure*}[!tp]
\centering
\includegraphics[width=0.94\textwidth,height=0.80\textheight,keepaspectratio]{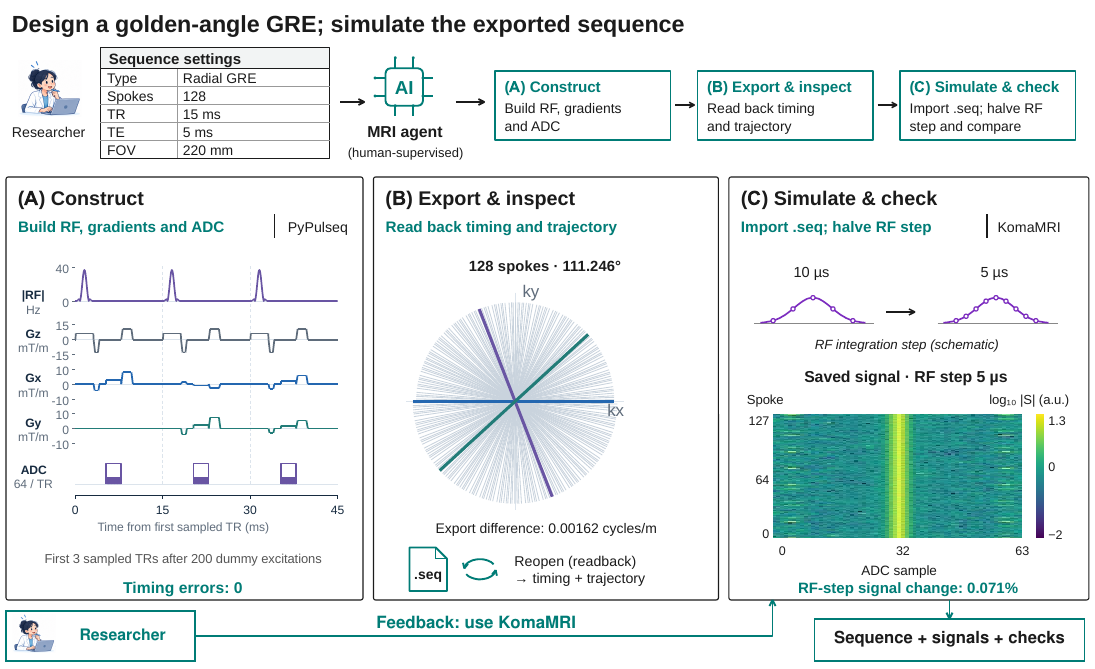}
\caption{Researcher-supervised sequence construction followed by a skill-guided audit. The setup table combines agreed settings and the recorded 128-spoke implementation. A: PyPulseq constructs RF, gradient and ADC events \cite{pypulseq,pulseq}. Native waveforms show the first three sampled repetitions after 200 dummy excitations; RF magnitude omits phase. Each repetition contains 64 ADC samples. Readback found no timing errors. B: Exported spokes rotate by 111.246 degrees \cite{winkelmann}. The unscaled geometry has nominal radius approximately 145.45 cycles/m; export/prewrite coordinates differ by at most 0.00162 cycles/m. C: At the researcher's request, the agent replaced an initial custom simulator with KomaMRI, inspected its installed API, ran its official EPI example and simulated the exported GRE \cite{koma}. RF-step curves are sampling schematics. The heatmap displays log10 magnitude of the saved 5-$\mu$s-step signal, with zero-based spoke/ADC indices and the stated display range. Halving the RF step from 10 to 5 $\mu$s changed the complete complex signal by 0.071\% (relative L2 change; exact value and definition in Supporting Information C.1 and Table S3). Action labels summarize records, not a transcript. Construction and simulation preceded the documented skill retrieval at audit time. These checks do not establish spatial convergence, image fidelity, SAR/PNS compliance or scanner readiness.}
\label{fig:3}
\end{figure*}

Two researcher-supervised Codex examples examined how the agent connected a request to tool use, checked outputs, and documented decisions (\hyperref[fig:3]{Figures~3}-\hyperref[fig:4]{4}). Saved scripts, reports, and project records document the requests, tool use, and revisions. Sequence guidance was retrieved during a later audit rather than original construction; reconstruction guidance was read during the run. The demonstrations follow these actions within individual projects, complementing the matched benchmark comparison. Supporting Information adds public task examples and retains numerical details, provenance, implementation history, and a separate diffusion example. NumPy, SciPy, and Matplotlib supported computation and display \cite{numpy,scipy,matplotlib}.

\subsubsection{Designing and checking a sequence}

The task was to design, export, and inspect a golden-angle radial gradient-echo (GRE) prototype. PyPulseq 1.5.0.post1 produced a Pulseq file with a 220-mm field of view, 5-mm slice, 10-degree nominal flip, TE/TR 5/15 ms, and 128 spokes of 64 samples \cite{pypulseq,pulseq}. Successive spokes rotated by 111.246 degrees \cite{winkelmann}. Configured per-axis gradient and slew limits were 32 mT/m and 130 T/m/s.

Sequence checks examined RF, gradient, and ADC events, timing, configured limits, and trajectory agreement before and after export. Readback tested whether the file preserved coordinates passed between applications. Simulation used KomaMRI on a 13,497-spin extruded 2D brain phantom \cite{koma}; its official EPI example served as an environment check. Halving the RF integration step tested temporal discretization on the same phantom. The workflow therefore checked both the exported file and the numerical signals produced from it.

\subsubsection{Comparing reconstructions}

The task was to reconstruct an acquired four-channel, 0.3-T T2-weighted M4Raw brain slice at two retrospective sampling settings \cite{m4raw,m4rawdata}. Source interpretation used HDF5 inspection with h5py and the release contract \cite{h5py,m4format}. The saved comparison comprised initial SENSE and subsequent L1-wavelet outputs, with shared ESPIRiT maps and L1 weight 0.01 at both masks \cite{sense,sigpy,sigpydocs,espirit,sparsemri,pywavelets}.

\begin{figure*}[!tp]
\centering
\includegraphics[width=0.94\textwidth,height=0.80\textheight,keepaspectratio]{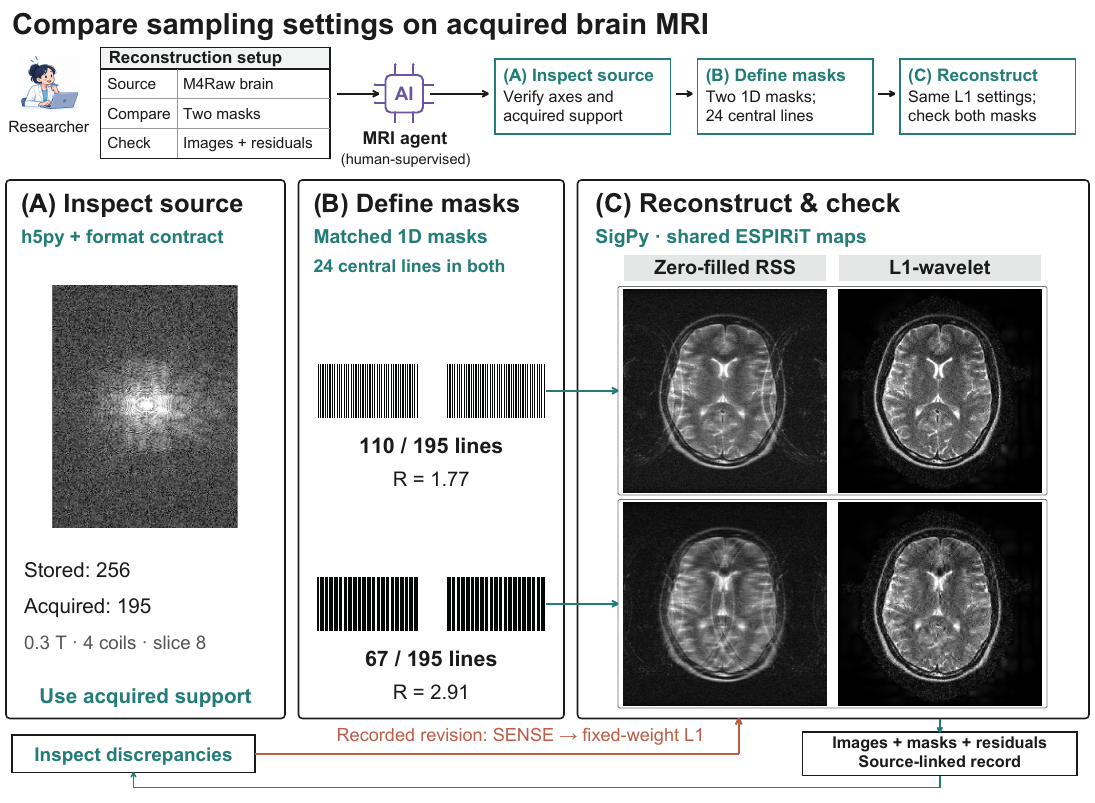}
\caption{Recorded reconstruction workflow on one four-coil, 0.3-T M4Raw T2-weighted brain slice \cite{m4raw,m4rawdata}. A: The release contract identifies the final array axis as phase encoding; the source has 195 acquired lines in a 256-column grid. Grayscale k-space shows log10 root-sum-of-squared coil magnitudes, with fixed bounds −1.306 to −0.223 (1st-99.5th percentiles of nonzero acquired values). K-space and mask displays are cropped to acquired support. B: Exact retrospective masks retain 110 or 67 acquired lines, including 24 central lines, giving effective accelerations of 1.77 and 2.91. Each mask connects to both reconstructions. C: Saved zero-filled RSS and L1-wavelet images share source-RSS normalization and a window of 0-0.676 of the source RSS maximum, without per-image rescaling or inferred anatomical flips. Both L1 conditions use shared ESPIRiT maps, db4 wavelets, weight 0.01 and 200 iterations \cite{sigpy,espirit,sparsemri}. The agent revised initial SENSE to fixed-weight L1 after inspecting disagreement. Zero-filled RSS remained closer to the noisy reference, and substantial data residuals remained unresolved. Supporting Information retains the initial results and complete diagnostics. Action labels summarize records, not a transcript; the setup table combines agreed settings and recorded choices. Brain-image pixels and the common window are unchanged. This supervised example is not a controlled agent-performance comparison.}
\label{fig:4}
\end{figure*}

The comparison preserved initial SENSE outputs and held sensitivity maps and reconstruction settings fixed between masks. We examined image discrepancy, data agreement, and numerical stability separately. Magnitude discrepancies used the noisy all-source root-sum-of-squares (RSS) image as reference. Sampled-line residuals assessed agreement with measurements used in the solve; withheld-line residuals used acquired measurements omitted from it. Additional iterations tested numerical stability. All images used the same source-derived normalization and display window. Together, these checks distinguish an optimization output that has stabilized from a reconstruction that agrees with the measured data.

\subsection{MRI Agent Bench and comparison}

\begin{figure*}[!tp]
\centering
\includegraphics[width=0.94\textwidth,height=0.80\textheight,keepaspectratio]{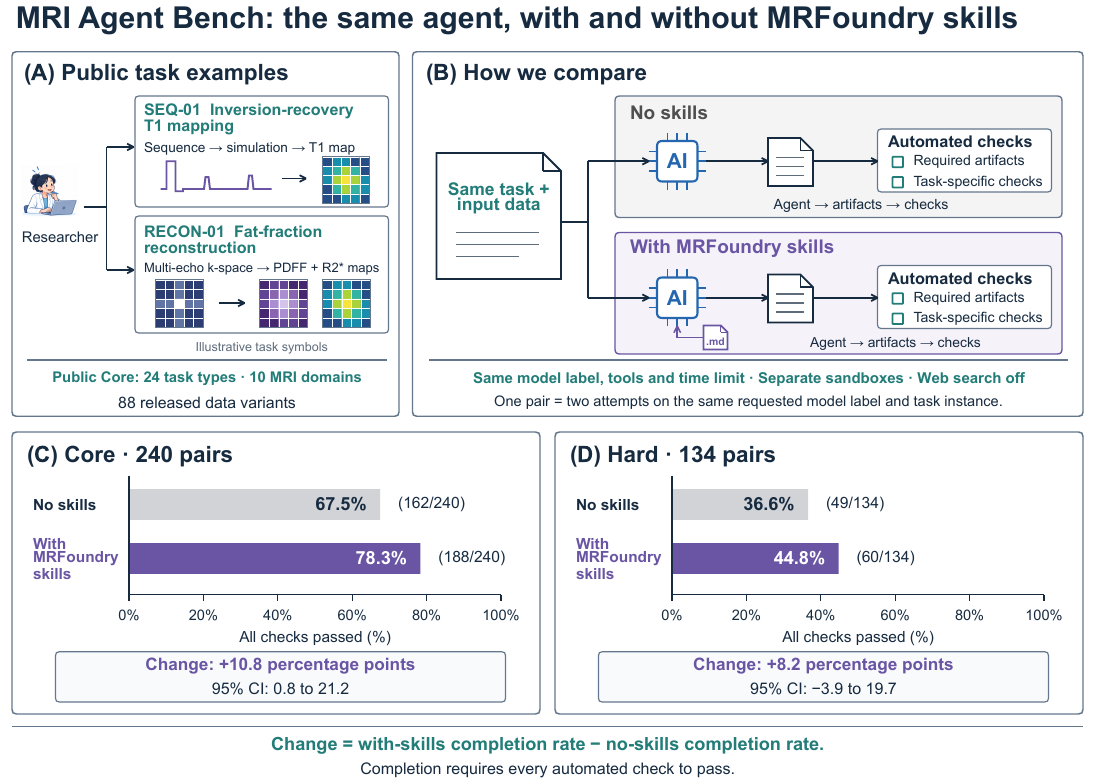}
\caption{MRI Agent Bench comparison. ``No skills'' excludes MRFoundry and its discovery instruction; common host system-skill catalogs were matched. A: public Core examples require inversion-recovery sequence simulation, reconstruction and T1 fitting (\href{https://github.com/KeWang0622/MRIAgentBench/blob/main/families/seq-01-inversion-recovery/task.example.json}{SEQ-01}), or PDFF/R2* reconstruction from multi-coil, multi-echo k-space (\href{https://github.com/KeWang0622/MRIAgentBench/blob/main/families/recon-01-fat-fraction/task.example.json}{RECON-01}) \cite{blochbenchtasks}. Icons are illustrative; these tasks differ from \hyperref[fig:3]{Figures~3}-\hyperref[fig:4]{4}. B: paired attempts use the same requested model label and task instance in separate sandboxes, with matched inputs, host tools and time limit. The intervention is the frozen MRFoundry package, including skills and helper scripts, plus its discovery instruction; skill reading is agent-selected. Web search is disabled, with package-index access allowed. Each arm produces artifacts assessed by the same checks; empty checklist symbols do not denote success. C-D: saved all-checks-passed rates on common 0-100\% scales. Change is with minus without skills, in percentage points; boxes report the saved two-way task/model cluster-bootstrap 95\% intervals. Core covers 24 task types $\times$ ten models (240 pairs). Hard combines three models $\times$ 15 types $\times$ two instances (90 pairs) and four models $\times$ 11 multi-stage types $\times$ one instance (44 pairs). Its interval includes zero, and unequal coverage precludes a common-task model ranking. Supporting Information gives resampling details, secondary unclustered tests and domain breakdowns.}
\label{fig:5}
\end{figure*}

\href{https://github.com/KeWang0622/MRIAgentBench}{MRI Agent Bench} complements the demonstrations with executable research tasks, specified inputs and artifacts, and automated checks (\hyperref[fig:5]{Figure~5}). Public Core contains 24 task types across ten MRI domains and 88 data variants \cite{blochbenchtasks,blochbench}. Coverage includes sequence design, reconstruction, quantitative imaging, analysis, and research planning and reporting. Hard comprises 11 multi-stage task types and four broader system-design types. A type specifies a problem structure; an instance fixes data and parameters. Detailed specifications and submission instructions are maintained in the accompanying repository. Separate author agents without access to the skills wrote specifications and private graders and generated data, parameters, and wording. Reference solutions and positive and negative controls checked instances before hashing and sealing.

We compared paired runs with and without MRFoundry. Each pair ran concurrently through Codex in separate sandboxes with the same requested model label, nominal reasoning setting, host tools, inputs, and time limit \cite{codex}. Search was disabled; network access was limited to package indexes. The intervention comprised the frozen MRFoundry package, including its skills, helper scripts, and project-level discovery instruction; the agent chose which guidance to read. The control omitted this package while retaining the same common host system-skill catalog. The snapshot was fixed before evaluated Core instances were opened; subsequently authored Hard tasks were not used to revise it \cite{mrfoundry014}.

Ten models each attempted one sealed instance per Core type, giving 240 pairs rather than evaluating all 88 variants. For Hard, three models attempted two instances of all 15 types (90 pairs); four further models attempted one instance of the 11 multi-stage types (44 pairs; \hyperref[tab:2]{Table~2}). Automated-check completion required every frozen check to pass. Checks examined submitted artifacts independently of the agent's report, including exported sequences, numerical outputs, code, and claims against supplied results. No language-model judge supplied completion labels.

\begin{table*}[!tp]
\caption{MRI Agent Bench comparison by model and evaluation cohort}
\label{tab:2}
\begingroup
\tablebodyfont
\setlength{\tabcolsep}{3.2pt}
\renewcommand{\arraystretch}{1.08}
\centering\begin{tabular}{@{}l>{\centering\arraybackslash}p{0.102\textwidth}>{\centering\arraybackslash}p{0.102\textwidth}>{\centering\arraybackslash}p{0.102\textwidth}>{\centering\arraybackslash}p{0.102\textwidth}>{\centering\arraybackslash}p{0.102\textwidth}>{\centering\arraybackslash}p{0.102\textwidth}@{}}
\toprule
 & \multicolumn{2}{c}{\textbf{Core: 24/model}} & \multicolumn{2}{c}{\textbf{Hard: 30/model}} & \multicolumn{2}{c}{\textbf{Hard: 11/model}}\\
\cmidrule(lr){2-3}\cmidrule(lr){4-5}\cmidrule(lr){6-7}
\textbf{Model} & {\tablecolheadfont \bfseries\shortstack{No skills}} & {\tablecolheadfont \bfseries\shortstack{With\\MRFoundry}} & {\tablecolheadfont \bfseries\shortstack{No skills}} & {\tablecolheadfont \bfseries\shortstack{With\\MRFoundry}} & {\tablecolheadfont \bfseries\shortstack{No skills}} & {\tablecolheadfont \bfseries\shortstack{With\\MRFoundry}}\\
\midrule
Claude Haiku 5.5 & 15 & \cellcolor{TableTint}19 & - & \cellcolor{TableTint}- & 2 & \cellcolor{TableTint}3\\
Claude Opus 5.5 & 22 & \cellcolor{TableTint}22 & 16 & \cellcolor{TableTint}20 & - & \cellcolor{TableTint}-\\
Claude Sonnet 5.5 & 17 & \cellcolor{TableTint}23 & - & \cellcolor{TableTint}- & - & \cellcolor{TableTint}-\\
Gemini 3.8 Flash & 19 & \cellcolor{TableTint}22 & - & \cellcolor{TableTint}- & 2 & \cellcolor{TableTint}4\\
Gemini 4 Argon & 21 & \cellcolor{TableTint}24 & 13 & \cellcolor{TableTint}14 & - & \cellcolor{TableTint}-\\
GPT-5.3 Codex & 9 & \cellcolor{TableTint}13 & - & \cellcolor{TableTint}- & 1 & \cellcolor{TableTint}1\\
GPT-6.1 Sol & 19 & \cellcolor{TableTint}22 & 15 & \cellcolor{TableTint}18 & - & \cellcolor{TableTint}-\\
GPT-6.1 Terra & 22 & \cellcolor{TableTint}22 & - & \cellcolor{TableTint}- & - & \cellcolor{TableTint}-\\
Kimi K3 & 13 & \cellcolor{TableTint}16 & - & \cellcolor{TableTint}- & - & \cellcolor{TableTint}-\\
MAI Code 1.1 Flash & 5 & \cellcolor{TableTint}5 & - & \cellcolor{TableTint}- & 0 & \cellcolor{TableTint}0\\
\midrule
\textbf{Completed} & \textbf{162} & \cellcolor{TableTint}\textbf{188} & \textbf{44} & \cellcolor{TableTint}\textbf{52} & \textbf{5} & \cellcolor{TableTint}\textbf{8}\\
\textbf{Paired trials} & \multicolumn{2}{c}{240} & \multicolumn{2}{c}{90} & \multicolumn{2}{c}{44}\\
\textbf{Discordant pairs} & \multicolumn{2}{c}{39 / 13} & \multicolumn{2}{c}{16 / 8} & \multicolumn{2}{c}{4 / 1}\\
\bottomrule\end{tabular}
\par\vspace{5pt}\raggedright\tablefootnotefont MRFoundry denotes the frozen evaluated snapshot (Supporting Information). ``No skills'' excludes MRFoundry and its discovery instruction; common host system-skill catalogs were matched. Completion requires all checks. Model names are recorded gateway request labels; effective backend snapshots and provider-side alias resolution were not independently retained. Labels are alphabetical; a dash denotes no evaluation. Core: one instance per type (24 types/model). Hard 30/model: two instances of 15 types; Hard 11/model: one instance of 11 multi-stage types. Discordant pairs: with-MRFoundry-only / control-only completions. Unequal Hard coverage precludes common-task model ranking. Pooled Hard: 49 to 60/134; system-design subset: 1 to 3/24. Figure 5 reports aggregate effect intervals.
\endgroup
\end{table*}

The primary effect was the difference in paired completion proportions. A two-way task-type/model cluster bootstrap supplied 95\% percentile intervals (4,000 resamples, seed 0), interpreted as sensitivity to the selected panel. Supporting Information gives the resampling algorithm, secondary unclustered McNemar comparisons, execution budgets, and grading limitations.

All runs that reached a model turn for the ten analyzed models were retained. Nine started pairs for an eleventh model were excluded after its API rejected the host's editing tool. We restarted 143 pair setups (94 Core, 49 Hard) that failed before either model turn, keeping task/model conditions unchanged. Memory and disk policies changed during evaluation; documented memory limits were matched within newly prepared pairs. Supporting Information records the policy changes and available execution information.

\section{Results}

The demonstrations connect researcher requests to sequence auditing, acquired-data reconstruction, and evidence-linked project records. The benchmark then measures task completion across models and MRI domains.

\subsection{Building and checking a sequence workflow}

For the sequence task, the agent produced a Pulseq file, trajectory plots, simulated signals, and readback checks (\hyperref[fig:3]{Figure~3}). The researcher asked the agent to replace its initial custom simulator with KomaMRI. The exported PyPulseq file became the simulation input, allowing the same sequence to be examined across the two tools.

During a later audit, the agent read MRFoundry sequence guidance, reopened the unchanged Pulseq file, and checked timing and gradient limits. It compared the readback trajectory with saved coordinates and inspected the saved simulation arrays. The audit report recorded the sequence hash, software environment, and check results. Timing and configured-limit checks passed; readback trajectories closely matched saved coordinates, and RF-step comparisons showed little signal change. The audit linked the exported sequence, its readback, and saved simulation outputs in one report. \hyperref[fig:3]{Figure~3} and Supporting Information retain the numerical checks.

\subsection{Inspecting and revising a reconstruction workflow}

For reconstruction, the agent assembled a comparison across sampling settings, including masks, sensitivity maps, complex images, and diagnostic reports (\hyperref[fig:4]{Figure~4}). After consulting guidance, the agent resolved array conventions and acquired support from the release contract. This interpretation determined which measurements could be retained and how acceleration was calculated. Fixed settings and a shared display window allowed the researcher to compare the two sampling conditions.

After inspecting the SENSE results, the agent tried L1-wavelet reconstruction. The records retain both attempts and their checks: the revision reduced magnitude discrepancy relative to SENSE, while zero-filled RSS remained closer to the noisy reference and measurement residuals remained unresolved. By saving the masks, settings, images, and diagnostic reports together, the agent made the change in method and its consequences available for researcher review. Supporting Information retains these diagnostics and the shared comparison conditions.

\subsection{Recording decisions for later work}

The agent recorded the acquired-support decision in project memory together with its supporting sources (\hyperref[fig:2]{Figure~2}). The entry identifies the source digest, release contract, and numerical report. Its reusable instruction is to verify axes and acquired support before masking; the source-specific dimensions remain attached to the original evidence. The record links the acceleration denominator to acquired measurements rather than the stored grid. It preserves the reasoning needed to revisit the decision alongside the commands and outputs.

\subsection{MRI Agent Bench comparison}

Across 240 Core pairs, runs with MRFoundry achieved 188 completions compared with 162 without it (78.3\% versus 67.5\%; difference 10.8 percentage points, 95\% CI 0.8 to 21.2). Of the discordant pairs, 39 completed only with MRFoundry and 13 only without it (\hyperref[fig:5]{Figure~5}). Seven models improved in aggregate Core count and three were unchanged (\hyperref[tab:2]{Table~2}). Benchmark exports retain the individual paired outcomes.

Across 134 Hard pairs, completion increased from 49 to 60 (36.6\% to 44.8\%; difference 8.2 points, 95\% CI −3.9 to 19.7), with 20 completions only with MRFoundry and nine only without it. \hyperref[tab:2]{Table~2} separates the Hard cohorts by task coverage. Supporting Information adds Core domain results, paired outcomes, and token-use measures. The observed completion rate was higher with the complete guidance package in both Core and Hard under the recorded execution conditions.

\section{Discussion}

\subsection{Connecting MRI knowledge to agent actions}

An MRI agent often needs to resolve data conventions or choose a validation check before it can use a numerical tool correctly. MRFoundry supplies instructions and sources for those decisions. Because the guidance is separate from both the host and numerical software, researchers can adopt it within an existing environment and inspect the instructions that shape the work. This is particularly useful for conventions that otherwise remain implicit, such as acquired support or the meaning of a reference image. An instruction can be traced to a source, revised when its assumptions change, and reviewed before wider use.

The paired comparison complements the workflow examples with higher aggregate completion on both Core and Hard when MRFoundry was available. Skills, references, helper scripts, and a discovery instruction are supplied together, and the agent selects relevant guidance. Researchers can inspect this material before a run and connect the resulting artifacts to the instructions and scientific sources that informed the work.

\subsection{Researcher supervision and memory reuse}

In the sequence example, the researcher directed the agent to an established simulator, and a later skill-guided audit connected the exported sequence to saved simulation arrays and checks. In reconstruction, the agent interpreted the source format, compared methods under shared settings, and retained the diagnostic reports. These examples show complementary roles: the researcher sets the question and directs revisions, while the agent carries out the computational steps and organizes the evidence needed to assess them.

Project memory carries the record of these decisions into a form that can be revisited in later sessions. Self-Refine studies revision through textual feedback, and Reflexion retains feedback across attempts \cite{selfrefine,reflexion}. MRFoundry records a lesson together with the evidence that defines its scope. The acquired-support example illustrates this recording step, including why a source-specific value must be rechecked. Researchers can inspect the observation, adapt the instruction, or propose a general contribution through review. The package therefore provides a practical structure for retaining local knowledge and refining shared MRI guidance.

\subsection{Future directions}

Repeated runs and broader panels are needed to narrow uncertainty in the Hard estimate and assess consistency across tasks and models. The Hard interval includes zero. Pair-level gains coexist with regressions; equal model totals can contain different successful tasks. Different Hard cohorts do not support a general model ranking.

Tasks were agent-authored within an author-selected scope; Core shares problem structure with development tasks, and independent expert review is incomplete. Automated checks assessed specified outputs rather than researcher effort, scientific judgment, or clinical validity. One Hard image subcheck accepted a zero-filled reconstruction on one instance; scores retain the original checks. Independent MRI review and stronger check validation are priorities. Core uses one run per task-model combination, leaving repeated-run reliability unmeasured. Nominal reasoning settings do not equalize provider budgets. Historical model snapshots, full configurations, and per-run resource exposure are unavailable. Setup incidents and shared resources limit interpretation; records do not establish uniform resources throughout.

The package comparison does not isolate individual skills, helper scripts, or retrieval decisions. The examples lack matched unguided baselines, and original sequence construction preceded the documented skill read. Longitudinal studies should test lesson retrieval, application, and reviewed contributions: the present study demonstrates recording but does not isolate performance benefits from memory reuse or community contributions.

Scientific validation should remain tied to the intended use. Sequence readback and RF-step sensitivity do not establish physical-model convergence or scanner readiness. In reconstruction, stable outputs retained unresolved measurement residuals, and the noisy RSS reference does not establish accuracy or consistently superior reconstruction. Independent references and stronger forward-model checks would extend the reconstruction validation. Research efficiency, scanner safety, and diagnostic benefit each require their own evaluation design.

\section{Conclusion}

MRFoundry connects MRI knowledge, established software, and researcher supervision through an extensible package of skills and evidence-linked project records. The demonstrations show how an agent can audit a sequence across tools, construct an acquired-data reconstruction comparison, and preserve the reasoning behind a research decision. Alongside the toolbox, MRI Agent Bench provides paired assessment across MRI task domains, with higher observed completion on both Core and Hard. MRFoundry gives researchers a practical way to inspect, adapt, and extend MRI workflow expertise within existing agent environments.

\section{Data and software availability}

MRFoundry source, contribution guidance, and tutorials are available at \url{https://github.com/KeWang0622/MRFoundry} \cite{mrfoundry,contributing,mrfoundrytutorial}, with the evaluated snapshot preserved \cite{mrfoundry014}. Ke Wang will maintain MRFoundry for at least five years after publication, providing support through GitHub issues and preserving the evaluated snapshot. MRI Agent Bench publishes Core tasks, data, checkers, and result exports \cite{blochbenchtasks,blochbench}. Hard materials, per-task outcomes, and traces remain private; public aggregates cannot reproduce its clustered interval. The \href{https://github.com/KeWang0622/MRFoundry/tree/f2a02ed482deb6a914140d5b85cbd0a77667a7ce/paper/evidence}{paper evidence directory} supplies original scripts, reports, input identities, and recorded environments, with portability gaps stated. Supplementary code provides a separate CPU-only basic-use example with pinned dependencies, synthetic input, export/readback checks, and expected outputs. Its clean-environment test does not repeat the historical agent evaluation. Reused human data remain in their original releases: M4Raw Brain v1.6 under CC BY 4.0 and Stanford HARDI under PDDL 1.0 \cite{m4rawdata,stanfordhardi}. Supporting Information describes the providers' approval, consent, and privacy conditions; no new participants were recruited.

\section{Funding}

No funding was received for this work.

\section{AI assistance disclosure}

Human authors led the study and manuscript, made scientific decisions, and interpreted the results. OpenAI Codex assisted software development, manuscript drafting, revision, and consistency checks. OpenAI's image-generation tool assisted explanatory character artwork and figure-layout concepts for \hyperref[fig:1]{Figures~1}-\hyperref[fig:5]{5} and S1; final layouts were assembled from editable components. Scientific MRI images and quantitative plots derive from recorded data and numerical outputs without generative enhancement. AI tools served as assistants; the authors retain responsibility for the scientific content and final manuscript.

\renewcommand{\bibfont}{\normalfont\fontsize{8.5bp}{10.5bp}\selectfont}
\bibliography{references}

\end{document}

\thispagestyle{fancy}
{\raggedright\hyphenpenalty=10000\rmfamily\Large\bfseries\color{black} MRFoundry: Knowledge-guided agent loops for end-to-end MRI research\par}
\vspace{4pt}
Ke Wang, Atakan Topcu, Hossein Momeni, and Molin Zhang\par
\vspace{4pt}

\section*{Supporting Information A}

\phantomsection\label{sec:A}

\addcontentsline{toc}{section}{Supporting Information A}

Additional research examples\par

\subsection*{A.1 From a researcher request to inspectable artifacts}

MRFoundry guidance connects a researcher-defined question with proposed actions, scientific tools, artifacts and checks (Figure S1). The four public task examples on page 2 illustrate this intended use; they are task-contract walkthroughs, not observed agent traces or newly executed demonstrations. The released outcome snapshot does not supply these models' generated artifacts or action logs. A separate recorded diffusion example follows on page 3. The researcher remains responsible for assumptions, interpretation and deciding whether unresolved checks require further work.\par

\begin{center}\includegraphics[width=\linewidth,height=0.70\textheight,keepaspectratio]{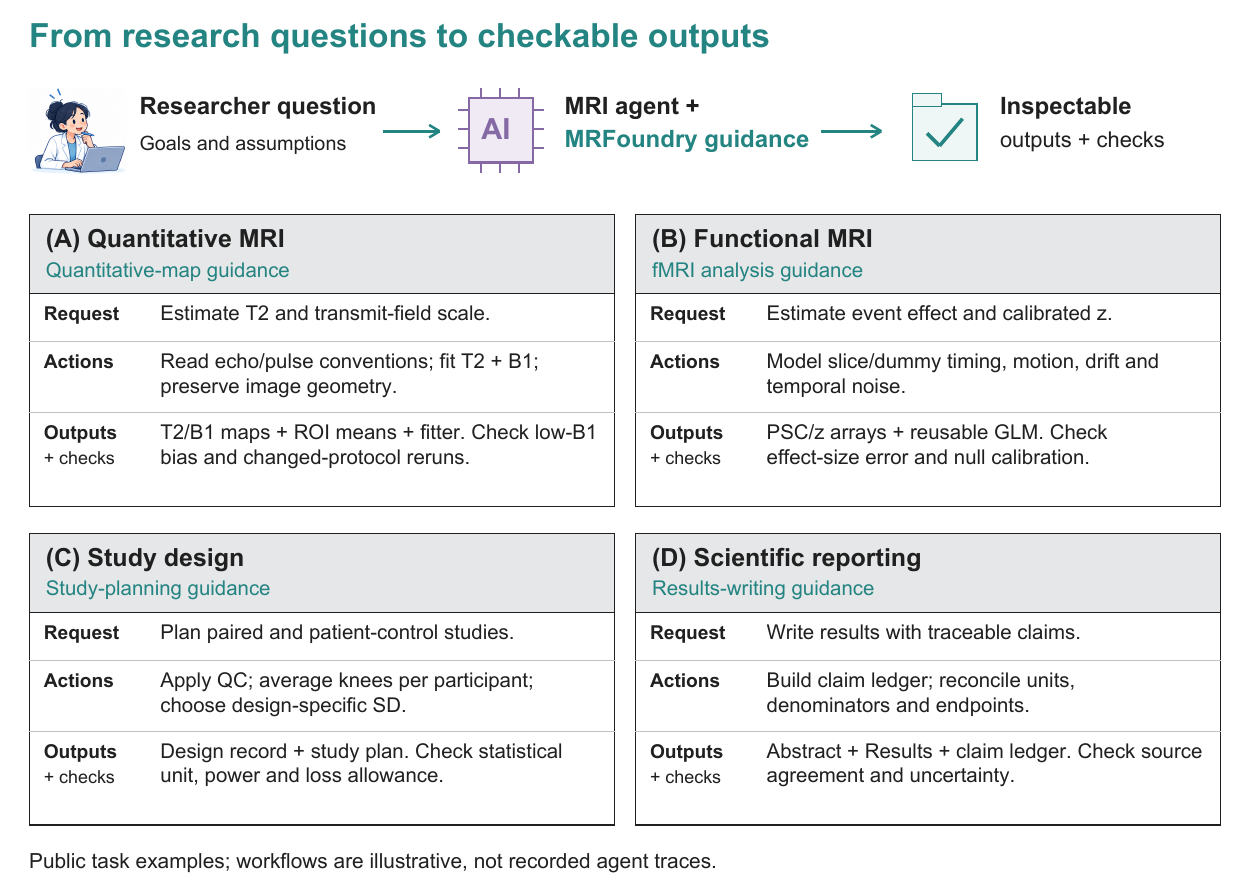}\end{center}

{\small Figure S1. Public task contracts illustrate how MRFoundry guidance can support additional research products. A, T2 and transmit-field mapping; B, event-related fMRI; C, pilot-based study design; D, traceable results reporting. Each panel links a request to suggested actions, required artifacts and checks. Arrows and guidance routing describe a possible workflow; they do not establish retrieval or tool use in a recorded trial. No generated MRI results are shown. Contracts and substantive checks are summarized in A.2; recorded paired outcomes are in Part B.\par}

Sources are pinned \href{https://github.com/KeWang0622/MRIAgentBench/tree/82d48e48685f6650648ba899ca90983121f90a30}{Core task contracts} (CC BY 4.0 descriptions/data; Apache-2.0 code) and \href{https://github.com/KeWang0622/MRFoundry/tree/bd49b9556146bcd70ab1026c52d67713f4cf2e7f}{MRFoundry guidance} (MIT). No model, grader or scientific workflow was run to prepare these examples.\par

\clearpage

\subsection*{A.2 Four public task-contract walkthroughs}

\textbf{T2 and transmit-field mapping.} A researcher requests true T2 and B1-scale maps from simulated multi-echo spin-echo data with imperfect refocusing. MRFoundry's \href{https://github.com/KeWang0622/MRFoundry/blob/bd49b9556146bcd70ab1026c52d67713f4cf2e7f/skills/mri-reconstruction/references/quantitative-maps.md}{quantitative-mapping instructions} explain stimulated-echo bias and prescribe an extended phase graph (EPG) dictionary with a bounded B1 grid. The proposed agent would check pulse conventions and timing, fit both parameters, and export \texttt{fit\_\allowbreak{}t2.py}, NIfTI maps and regional summaries. The researcher could then inspect the fitted B1 range and judge model adequacy using the requirement for absolute lowest-B1-quartile median T2 bias ≤3\%; saved single-exponential controls fail this check. The \href{https://github.com/KeWang0622/MRIAgentBench/tree/82d48e48685f6650648ba899ca90983121f90a30/families/v4-t2epg}{task contract and checks} give the remaining outputs, tolerances and changed-acquisition checks.\par

\textbf{Event-related fMRI.} A researcher requests single-event percent-signal-change (PSC) and calibrated positive-response z maps from synthetic BOLD data. MRFoundry's \href{https://github.com/KeWang0622/MRFoundry/blob/bd49b9556146bcd70ab1026c52d67713f4cf2e7f/skills/mri-research/references/analysis-processing.md}{MRI analysis guide} directs agents to established GLM tools and asks them to report the statistical unit and pipeline versions. For this contract, the proposed agent would align event, slice and dummy-volume timing, model motion, drift and temporal noise, and produce \texttt{fmri\_\allowbreak{}glm.py}, PSC/z arrays, a figure and a report. Before interpreting activation, the researcher would inspect null calibration: the required null-tail fraction at z>1.645 is [0.02,0.085]; saved white-noise OLS controls fail calibration. The \href{https://github.com/KeWang0622/MRIAgentBench/tree/82d48e48685f6650648ba899ca90983121f90a30/families/v7-fmriglm}{task contract and checks} supply the remaining accuracy, uncertainty and rerun requirements.\par

\textbf{Pilot-based study design.} A researcher requests recruitment plans for a paired cartilage-T2 loading challenge and a patient-control comparison. MRFoundry's \href{https://github.com/KeWang0622/MRFoundry/blob/bd49b9556146bcd70ab1026c52d67713f4cf2e7f/skills/mri-research-workflow/references/plan-sample-size.md}{sample-size instructions} require QC before forming participant-level endpoints and distinguish paired-difference from between-participant variance. The proposed agent would average knees within participant/session, estimate repeatability, select the appropriate variance and inflate for nonevaluable participants. The resulting \texttt{design.json} and plan of at most 1,000 words would expose assumptions for the researcher to review before choosing a design. One informative check rejects counting knees as independent participants. The \href{https://github.com/KeWang0622/MRIAgentBench/tree/82d48e48685f6650648ba899ca90983121f90a30/families/v6-pilotdesign}{task contract and checks} supply effect, power and loss assumptions; these synthetic values require reconsideration for a new study.\par

\textbf{Traceable results reporting.} A researcher requests an abstract, Results section and claim ledger from a fictional MRI trial. MRFoundry's \href{https://github.com/KeWang0622/MRFoundry/blob/bd49b9556146bcd70ab1026c52d67713f4cf2e7f/skills/mri-research-workflow/references/write-results.md}{results-writing instructions} require building the ledger before drafting, retaining prespecified endpoints regardless of significance, and converting units in code. The proposed agent would link estimates to source entries, then write \texttt{abstract.md}, \texttt{results.md} and \texttt{claims.json}. Point estimates, intervals and denominators would remain linked to their inputs. The researcher could audit each sentence against its ledger before approving the interpretation. One source-agreement check is that E04 retains its nonsignificant result after −0.0182 s becomes −18.2 ms, with Holm p=0.12 unchanged. This illustrates a check on fictional input; the \href{https://github.com/KeWang0622/MRIAgentBench/tree/82d48e48685f6650648ba899ca90983121f90a30/families/v6-honestresults}{task contract and checks} give the remaining requirements.\par

These routes are proposed uses, not recorded guidance retrieval or helper execution. The public catalogue covers 24 task types and 88 variants; the frozen comparison evaluated one sealed instance per type, not all variants.\par

\clearpage

\subsection*{A.3 Recorded diffusion example: maps and same-data fit checks}

Stanford HARDI supplied one acquired subject (81×106×76 voxels, 2-mm isotropic; 10 b=0 and 150 b=2000 s/mm$^2$ volumes). DIPY 1.12.1 fitted weighted-least-squares tensors on fixed slices 30, 38 and 46; S0 was mean b=0. A three-dimensional median\_otsu mask used radius 3, one pass and dilation 1, with holes filled per displayed slice. This is independent of the sequence and reconstruction demonstrations. Figure S2 preserves the recorded maps and residuals; no fitting or preprocessing was rerun.\par

\begin{center}\includegraphics[width=\linewidth,height=0.60\textheight,keepaspectratio]{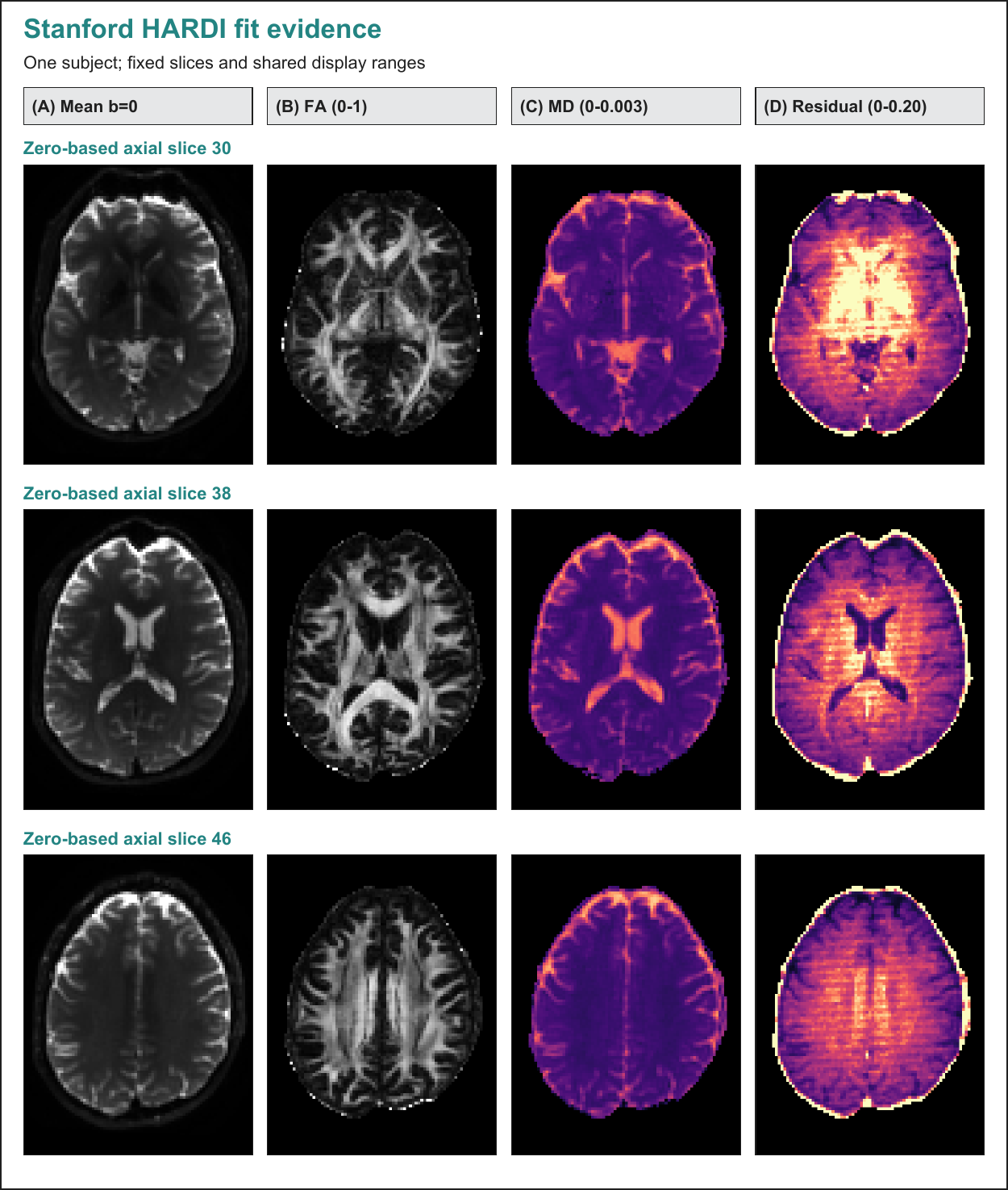}\end{center}

{\small Figure S2. Nonzero same-data residuals remain across three slices of one Stanford HARDI subject. Rows show zero-based slices 30, 38 and 46; columns show (A) mean b=0, (B) FA, (C) MD and (D) signal-fit residual. Original PNG pixels and shared windows are retained: mean b=0, 0-3989 source units (rounded); FA, 0-1; MD, 0-0.003 mm$^2$/s; residual, 0-0.20. Screen left is L, right is R and top is anterior; fitted maps are black outside the mask. Mask counts are 4,799, 4,834 and 4,365; median residuals are 10.24\%, 8.95\% and 9.05\%. Residual is $L_2$(predicted - measured) / max($L_2$(measured), 1) across all 160 volumes; the floor is inactive. These are within-subject examples, not independent subjects. Source: Rokem et al., Stanford HARDI (PDDL 1.0).\par}

Provider motion correction was not independently verified or repeated; susceptibility/eddy correction and anatomical gradient-frame alignment were not validated. FA checks followed numpy.nan\_to\_num, so raw FA finiteness was not established. The b=2000 protocol is not optimal conventional DTI. The residual describes model agreement, not biological accuracy or a clinical threshold. \href{https://github.com/KeWang0622/MRFoundry/tree/f2a02ed482deb6a914140d5b85cbd0a77667a7ce/paper/evidence/demonstrations/acquired}{Source and retained fit evidence}.\par

\clearpage

\section*{Supporting Information B}

\phantomsection\label{sec:B}

\addcontentsline{toc}{section}{Supporting Information B}

Additional evaluation results\par

The frozen Core panel contains 240 pairs (24 task types × ten requested model labels); Hard contains 134 pairs with unequal model/type coverage. A completed attempt passes every automated check. Table S1 preserves paired effects; Table S2 includes domains with gains, ties and losses. These are author-selected, largely synthetic tasks; independent external MRI-expert review is pending. Model labels are recorded gateway requests, not independently verified backend snapshots. No model, grader, bootstrap or token analysis was rerun.\par

\subsection*{Table S1. Aggregate completion and paired effects}

\begingroup\fontsize{10}{11.5}\selectfont\setlength{\tabcolsep}{5pt}
\begin{longtable}{@{}>{\raggedright\arraybackslash}p{\dimexpr0.09\linewidth-2\tabcolsep\relax}>{\raggedright\arraybackslash}p{\dimexpr0.08\linewidth-2\tabcolsep\relax}>{\raggedright\arraybackslash}p{\dimexpr0.26\linewidth-2\tabcolsep\relax}>{\raggedright\arraybackslash}p{\dimexpr0.14\linewidth-2\tabcolsep\relax}>{\raggedright\arraybackslash}p{\dimexpr0.24\linewidth-2\tabcolsep\relax}>{\raggedright\arraybackslash}p{\dimexpr0.19\linewidth-2\tabcolsep\relax}@{}}
\toprule\rowcolor{HeaderGrey}
\textbf{Set} & \textbf{Pairs} & \textbf{Without $\rightarrow$ with} & \textbf{Change, pp} & \textbf{95\% CI, pp} & \textbf{P}\\
\midrule\endfirsthead
\toprule\rowcolor{HeaderGrey}
\textbf{Set} & \textbf{Pairs} & \textbf{Without $\rightarrow$ with} & \textbf{Change, pp} & \textbf{95\% CI, pp} & \textbf{P}\\
\midrule\endhead
\bottomrule\endfoot
Core & 240 & 162/240 $\rightarrow$ 188/240 & +10.8 & [+0.8, +21.2] & 0.0004\\
Hard & 134 & 49/134 $\rightarrow$ 60/134 & +8.2 & [-3.9, +19.7] & 0.061\\
\end{longtable}\endgroup

\subsection*{Table S2. Core completion by domain}

\begingroup\fontsize{10}{11.5}\selectfont\setlength{\tabcolsep}{5pt}
\begin{longtable}{@{}>{\raggedright\arraybackslash}p{\dimexpr0.31\linewidth-2\tabcolsep\relax}>{\raggedright\arraybackslash}p{\dimexpr0.1\linewidth-2\tabcolsep\relax}>{\raggedright\arraybackslash}p{\dimexpr0.28\linewidth-2\tabcolsep\relax}>{\raggedright\arraybackslash}p{\dimexpr0.31\linewidth-2\tabcolsep\relax}@{}}
\toprule\rowcolor{HeaderGrey}
\textbf{Domain} & \textbf{Pairs} & \textbf{Without $\rightarrow$ with} & \textbf{With-only / without-only}\\
\midrule\endfirsthead
\toprule\rowcolor{HeaderGrey}
\textbf{Domain} & \textbf{Pairs} & \textbf{Without $\rightarrow$ with} & \textbf{With-only / without-only}\\
\midrule\endhead
\bottomrule\endfoot
Diffusion & 20 & 17/20 $\rightarrow$ 15/20 & 0 / 2\\
Functional MRI & 10 & 8/10 $\rightarrow$ 9/10 & 1 / 0\\
Hardware / field & 20 & 13/20 $\rightarrow$ 14/20 & 3 / 2\\
Perfusion / CEST & 20 & 14/20 $\rightarrow$ 18/20 & 4 / 0\\
Reconstruction & 20 & 13/20 $\rightarrow$ 15/20 & 4 / 2\\
Relaxometry / QSM & 20 & 13/20 $\rightarrow$ 19/20 & 6 / 0\\
Sequence design & 70 & 34/70 $\rightarrow$ 45/70 & 14 / 3\\
Spectroscopy & 10 & 8/10 $\rightarrow$ 8/10 & 1 / 1\\
Study design / statistics & 20 & 17/20 $\rightarrow$ 17/20 & 2 / 2\\
Writing / review & 30 & 25/30 $\rightarrow$ 28/30 & 4 / 1\\
\end{longtable}\endgroup

\textbf{Uncertainty and paired comparisons.} The saved two-way cluster bootstrap independently resamples task-type and model labels with replacement, weights each observed cell by the product of label multiplicities, includes its retained instances together, and averages paired differences. Missing cells contribute no observations. Each set uses 4,000 draws with Python random.Random(0); sorted nonempty means supply indices int(0.025N) and int(0.975N)-1. Thus Core weights types rather than domains; Hard preserves unequal instance/cohort coverage. The Hard interval includes zero. Exact two-sided McNemar P values (Core 0.0004; Hard 0.061) are secondary, unclustered comparisons assuming independent discordant-pair signs. Discordances are 39 with-only/13 without-only for Core and 20/9 for Hard; no multiplicity correction is claimed.\par

\textbf{Interpretation and accounting.} Diffusion declined from 17/20 to 15/20, with no gains; the failed subchecks are not identified and no instruction-level cause is inferred. Hard system-design tasks remained largely unsolved (1/24 versus 3/24). Total-token with/without ratios were 1.048 (Core) and 1.062 (Hard); tokens per completed attempt ratios were 0.903 and 0.868, counting tokens from unsuccessful attempts. These are reported token measures, not money, latency or energy; non-OpenAI reasoning-token accounting was incomplete. One attempt per combination does not establish repeated-run reliability or a general model ranking. Public Hard aggregates cannot recover the task/model cells needed to recompute its interval.\par

{\small\raggedright Sources: \href{https://github.com/KeWang0622/MRIAgentBench/blob/82d48e48685f6650648ba899ca90983121f90a30/leaderboard/offline-results-2026-11-03.json}{frozen public Core results}, \href{https://github.com/KeWang0622/MRFoundry/tree/f2a02ed482deb6a914140d5b85cbd0a77667a7ce/paper/evidence/evaluation}{original statistics helper and provenance}. Hard cohort coverage remains in main Table 2; execution and reproduction limits are summarized in C.3.\par}

\clearpage

\section*{Supporting Information C}

\phantomsection\label{sec:C}

\addcontentsline{toc}{section}{Supporting Information C}

Implementation, reproducibility and public resources\par

\subsection*{C.1 Recorded sequence construction and later audit}

PyPulseq 1.5.0.post1 exported and reopened a golden-angle radial GRE. Researcher feedback replaced a custom simulator with KomaMRI, which simulated 328 excitations on 13,497 spins (409 in-plane locations × 33 z positions). Table S3 retains the checks. Construction/simulation occurred on 23 September 2026; the skill read was a later 28 October audit of unchanged artifacts, not established original retrieval. RF-step agreement tests temporal sensitivity on this phantom, not spatial convergence or scanner/SAR/PNS safety.\par

\subsection*{Table S3. Sequence readback and simulation checks}

\begingroup\fontsize{10}{11.5}\selectfont\setlength{\tabcolsep}{5pt}
\begin{longtable}{@{}>{\raggedright\arraybackslash}p{\dimexpr0.46\linewidth-2\tabcolsep\relax}>{\raggedright\arraybackslash}p{\dimexpr0.22\linewidth-2\tabcolsep\relax}>{\raggedright\arraybackslash}p{\dimexpr0.32\linewidth-2\tabcolsep\relax}@{}}
\toprule\rowcolor{HeaderGrey}
\textbf{Recorded check} & \textbf{Value} & \textbf{Reference or scope}\\
\midrule\endfirsthead
\toprule\rowcolor{HeaderGrey}
\textbf{Recorded check} & \textbf{Value} & \textbf{Reference or scope}\\
\midrule\endhead
\bottomrule\endfoot
TE / TR & 5.000 / 15.000 ms & Readback; target 5 / 15 ms\\
Timing errors & 0 & PyPulseq readback\\
Peak per-axis gradient & 12.735 mT/m & Configured limit 32 mT/m\\
Peak per-axis slew & 125.266 T/m/s & Configured limit 130 T/m/s\\
Export/prewrite trajectory difference & 0.00162 cycles/m & Maximum absolute coordinate difference\\
Retained ADC samples & 8,192 & Finite and nonzero; 128 x 64\\
RF-step signal change & 0.07096\% & $L_2$(S5us - S10us) / $L_2$(S5us)\\
\end{longtable}\endgroup

\subsection*{C.2 Recorded reconstruction revision and unresolved agreement}

One M4Raw Brain v1.6 slice (four channels, 0.3 T) had 195 acquired PE lines within a 256-column grid. Masks intersected nonzero stored indices 30-224 with indices divisible by 2 or 4, plus 116-139, retaining 110/67 lines (R 1.77/2.91); both shared 24-line ESPIRiT calibration. Initial L2-SENSE used weight 0.001/120 iterations. The bounded revision retained the maps and used fixed L1-wavelet weight 0.01, db4, 200 iterations and seed 0 for both masks, without a per-mask search. All outputs share source-RSS normalization; no image gain was fitted.\par

\subsection*{Table S4. Recorded reconstruction comparisons}

\begingroup\fontsize{10}{11.5}\selectfont\setlength{\tabcolsep}{5pt}
\begin{longtable}{@{}>{\raggedright\arraybackslash}p{\dimexpr0.6\linewidth-2\tabcolsep\relax}>{\raggedright\arraybackslash}p{\dimexpr0.2\linewidth-2\tabcolsep\relax}>{\raggedright\arraybackslash}p{\dimexpr0.2\linewidth-2\tabcolsep\relax}@{}}
\toprule\rowcolor{HeaderGrey}
\textbf{Saved metric (\%)} & \textbf{110 lines} & \textbf{67 lines}\\
\midrule\endfirsthead
\toprule\rowcolor{HeaderGrey}
\textbf{Saved metric (\%)} & \textbf{110 lines} & \textbf{67 lines}\\
\midrule\endhead
\bottomrule\endfoot
Initial SENSE magnitude discrepancy & 43.55 & 70.26\\
Final L1 magnitude discrepancy & 30.33 & 33.62\\
Zero-filled RSS magnitude discrepancy & 21.24 & 28.92\\
Initial SENSE sampled-data residual & 30.08 & 21.87\\
Final L1 sampled-data residual & 30.88 & 24.84\\
Initial SENSE withheld-source residual & 103.11 & 202.50\\
Final L1 withheld-source residual & 85.72 & 91.38\\
Final L1 200-to-400-iteration change & 0.0477 & 0.0741\\
\end{longtable}\endgroup

Magnitude discrepancy is full-matrix $L_2$(|x| - RSS) / $L_2$(RSS). Data residuals normalize complex k-space disagreement by observed-data norm on sampled or withheld masks; withheld lines do not enter the solve. Iteration change is $L_2$(x400 - x200) / $L_2$(x400). L1 reduced discrepancy versus initial SENSE, but zero-filled RSS remained closer to the noisy source and sampled residual increased. Stability does not establish anatomical accuracy; noise covariance/prewhitening and biological validity were not assessed. The saved acquired-support lesson was not tested in a later task. \href{https://github.com/KeWang0622/MRFoundry/tree/f2a02ed482deb6a914140d5b85cbd0a77667a7ce/paper/evidence/demonstrations}{Scripts, initial outputs and reports}.\par

\clearpage

\subsection*{C.3 Execution conditions and practical reproduction}

To use the extensible package, follow the \href{https://github.com/KeWang0622/MRFoundry\#quick-start}{repository Quick Start} with a compatible coding agent and install the required scientific software separately. The initial guidance collection is a starting point for further contributions.\par

Paired arms used Codex CLI 0.159.2, the same requested model label, nominal medium reasoning, tools, inputs and within-pair time limit, with separate Linux namespaces, Python environments and Julia depots. Web search was disabled; package-index access was allowed. The control lacked MRFoundry and its discovery instruction. Core caps were 45 min for two relaxometry/QSM types, three writing/review types and pilot design, and 60 min otherwise. Hard caps were 120/150/180/210 min for 3/6/2/4 types. Nominal reasoning settings did not equalize vendor reasoning budgets. All model-reaching trials for the ten analyzed models were retained; nine started pairs for an eleventh model were excluded after API rejection of file editing. The 143 pre-turn setup failures (94 Core, 49 Hard) were restarted unchanged.\par

Memory and disk policies changed during evaluation. An 8-GiB process cap was documented as matched within newly prepared pairs; a 60-GiB workspace/temporary-storage guard followed a retained run that generated 232 GB of temporary files. Complete exposure/transition and arm-specific guard-termination records are unavailable; uniform resources and equivalent historical disk exposure cannot be established. One Hard instance accepted a zero-filled image subcheck; this did not alone imply whole-task completion, and the recorded outcome was retained. \href{https://github.com/KeWang0622/MRFoundry/blob/f2a02ed482deb6a914140d5b85cbd0a77667a7ce/paper/evidence/evaluation/HISTORICAL-EXECUTION.md}{Historical execution manifest}.\par

The \href{https://github.com/KeWang0622/MRFoundry/blob/f2a02ed482deb6a914140d5b85cbd0a77667a7ce/paper/evidence/README.md}{evidence archive} preserves scripts, reports and hashes; local paths/private run-note dependencies require adaptation, and large derived arrays are omitted. Checksum identity is not numerical reproduction. The separate Code Review example provides a hash-locked Python 3.12/PyPulseq installation and an 8×8 Cartesian GRE helper-use test. Its recorded TE/TR are 6/20 ms, with 64 ADC samples, 18 exported-file checks, read-back timing and analytic-grid agreement; a wrong 7-ms TE expectation is rejected. This new CPU-only basic-use test does not rerun historical agent tasks, simulations or MRI reconstructions, and does not establish scanner safety.\par

\subsection*{C.4 Secondary human-data provenance}

The \href{https://huggingface.co/datasets/mylyu/M4Raw_brain/blob/v1.6.0/DATA_CARD.md}{M4Raw v1.6 data card} reports original Shenzhen Technology University IRB approval SZTULL012021005 and written consent including anonymized public release (CC BY 4.0). The \href{https://purl.stanford.edu/yx282xq2090}{Stanford HARDI record} supplies PDDL 1.0 and privacy/no-identification conditions, but the checked record does not specify original ethics approval or public-release consent; neither is inferred. These are source-attributed conditions; no local approval, exemption or waiver is claimed.\par

\subsection*{C.5 Contribution guidance and online demonstrations}

The \href{https://kewang0622.github.io/slides/mrfoundry/}{interactive teaching tutorial} uses client-side analytic/synthetic MRI figures and controls; it does not host agent inference. The \href{https://kewang0622.github.io/mrfoundry/examples/}{workflow examples} replay saved sequence, simulation and reconstruction artifacts with explanatory controls; they are prerecorded presentations, not live agent runs. \href{https://github.com/KeWang0622/MRFoundry/blob/main/CONTRIBUTING.md}{Contribution guidance} asks for primary sources, a pull request and changelog, automated checks and maintainer review. Project memory should retain evidence, scope and revalidation conditions; private context should be removed before public contribution. Later memory reuse and contribution-driven performance gains were not evaluated here.\par